\documentclass[useAMS,usenatbib,referee]{mn2e}
\usepackage{graphicx}
\usepackage{subfigure}
\usepackage{latexsym}
\def\prodi#1#2{\frac{{\rm d} #1}{{\rm d} #2}}

\def\parti#1#2{\frac{\partial #1}{\partial #2}}

\newcommand\araa{{ARA\&A}}%
\newcommand\apj{{ApJ}}%
\newcommand\apjl{{ApJ}}%
\newcommand\apss{{Ap\&SS}}%
\newcommand\aap{{A\&A}}%
\newcommand\mnras{{MNRAS}}%
\begin{document}

\title[Interaction of winds in binary system
PSR~1259-63/LS2883. ]{Modelling interaction of relativistic and
  non-relativistic winds in binary system PSR
  B1259-63/SS2883\footnote{Star SS2883 in what follows will be
    refereed as LS2883 according to the standard catalog of Luminous Stars in
    the Southern Milky Way \citep{stephenson71}} - II.~Impact of magnetization and
  anisotropy of the pulsar wind}

\author[Bogovalov et. al.]
{Bogovalov${}^1$ S.V.,Khangulyan${}^2$ D.,Koldoba${}^3$ A.V.,Ustyugova${}^3$ G.V.,Aharonian${}^{4,5}$ F.A.\\
  ${}^1$Moscow Engineering Physics Institute (state university), Moscow, Russia,\\
  ${}^2$Institute of Space and Astronautical Science/JAXA, Sagamihara, Japan,\\
  ${}^3$Keldysh Institute of Applied Mathematics RAS, Moscow, Russia,\\
  ${}^4$Dublin Institute for Advanced Studies, Dublin, Ireland,\\
  ${}^5$Max-Planck-Institut f\"ur Kernphysik, Heidelberg, Germany.}
\date{ } 

\maketitle

\begin{abstract}
  In this paper, we present a numerical study of the properties of the
  flow produced by the collision of a magnetized anisotropic pulsar wind
  with its environment in binary system. We compare the impact of both the
  magnetic field and the wind anisotropy to the benchmark case of a
  purely hydrodynamical (HD) interaction of isotropic winds, which has
  been studied in detail by \citet{bogovalov08}.  We consider the
  interaction in axisymmetric approximation, i.e. the pulsar rotation
  axis is assumed to be oriented along the line between the pulsar and
  the optical star and the effects related to the pulsar orbiting are
  neglected. The impact of the magnetic field is studied for the case
  of weak magnetization (with magnetization parameter $\sigma<0.1$),
  which is consistent with  conventional models of pulsar winds. The
  effects related to anisotropy in pulsar winds are modeled
  assuming that the kinetic energy flux in a non-magnetized pulsar wind
  is strongly anisotropic, with the minimum at the pulsar rotation axis
  and the maximum in the perpendicular direction.  We show that,
  although both considered effects change the shape of the region
  occupied by the terminated pulsar wind, their impact appears to be
  small. In particular, for the magnetization of the pulsar wind below
  0.1, the magnetic field pressure remains well below the plasma pressure
  in the post-shock region. Thus, in the case of interaction of a pulsar
  with the stellar wind environment (opposite to the case of plerions,
  i.e. the pulsar interaction with the interstellar medium, when the magnetic
  field becomes dynamically important independently on the wind
  magnetization) the HD approach represents a feasible approximation
  for numerical modelling.
\end{abstract}

\section{Introduction}
Pulsars lose their rotation energy through relativistic winds, the
interaction of which with the surrounding environment  leads to the
formation of so-called Pulsar Wind Nebulae (PWNe).  PWNs are bright in
X- and gamma ray regions filled with ultrarelativistic electrons
accelerated at the pulsar wind termination shock
\citep{rees74,kennel84}. While Crab Nebula is the most famous
(although rather atypical) representative, the recent X-ray and TeV
gamma ray observations have shown \citep[see e.g.][]{gaensler06} that
the formation of PNWe is a common phenomenon for many other pulsars as
well. Importantly, the physical conditions, and consequently their
observational properties, appear to differ significantly in different
PWNs. A very interesting situation arises when a pulsar is located in
a binary system. In this case the pulsar wind interacts with the wind
from the companion star.  This case, in particular, is realized in the
binary system PSR1259-63/LS2883 which consists of a $\sim 47.8 \rm ms$
pulsar in an elliptic orbit, with eccentricity $e=0.87$, around a massive
fast rotating O8 optical companion \citep{johnston92,negueruela11}. Observations
show that this system is a strongly variable source of non-thermal
emission extending from radio to TeV gamma-rays \citep[see e.g.][and
references therein]{neronov07}.

Two other variable TeV galactic gamma-ray sources, LS 5039 and
LS~I~+61~303 \citep[see e.g.][]{paredes00,bosch-ramon09}, are discussed
as possible, although less evident candidates for being a part of a  ``binary
pulsar'' source population \citep{maraschi81,martocchia05}.  LS 5039
consists of an O6.5V star and an unidentified compact object in a 3.9 days
orbit.  This object has been detected as a periodic source of gamma
rays by HESS \citep{aharonian06,aharonian05*c} and by {\it Fermi}
Large Area Telescope (LAT) \citep{abdo09*b}. In LS~I~+61~303 an
unidentified compact source moves around a B0Ve star in an eccentric
orbit with a period of 26.5 days.  This source was as well detected in
GeV and TeV gamma rays
\citep{albert06,abdo09,anderhub09,acciari11}. Although, the compact
source nature in these objects (black hole or neutron star) is not
yet firmly established, the binary pulsar scenario provides a possible
explanation of the broadband non-thermal emission detected from these
sources \citep[see e.g.][]{dubus06*b}.

The collision of supersonic winds from two stars located in a binary
system results in formation of two terminating shock fronts and a
tangential discontinuity separating relativistic and nonrelativistic
parts of the shocked flow. It has been found from the detailed
hydrodynamical modelling \citep{bogovalov08} that the shocked flow
propagates into a limited solid angle with rather high velocity. It
has been shown that on the scales, which characterize the binary
system, the shocked pulsar wind material obtains relativistic bulk
velocities. As proposed by \citet{khangulyan08}, this should lead to
the formation of strongly variable non-thermal emission due to the
Doppler boosting, thus a detailed study of the properties of the
post-shock flow is important for proper interpretation of variable X-
and gamma rays detected from the system
\citep{aharonian05,chernyakova06,uchiyama09,aharonian09,abdo11}.


In the previous pure HD study by \citet{bogovalov08} several
potentially important effects have been identified for the future
research. In particular, the development of Kelvin-Helmholtz
instability at the contact discontinuity and orbital motion of the
pulsar should enhance the mixing of pulsar and stellar wind materials
on a spacial scales slightly exceeding the size of the binary system
\citep{bosch-ramon11}. Given relatively short cooling time of VHE
electrons \citep[see e.g.][]{khangulyan07} this mixing should not
significantly affect the variable gamma ray component, which is
expected to be produced inside the system. On the other hand, 
several other effects related to impact of magnetic field and
anisotropy of the pulsar wind, which are almost unavoidable features
of the pulsar wind, appear to be important already on the scales of
binary system.  The importance of this effects can be illustrated by
the impact of the magnetic field on the interaction of the pulsar wind
with interstellar medium. Even a weak magnetization of the pulsar wind
can result in a dramatic modification of the flow in the post shock
region \citep{komissarov03,bogovalov05,bucciantini05}. In this paper
we present a study of the role of the magnetic field and anisotropic
energy flux on the dynamic of the shocked material in binary pulsar
systems.

\section{Properties of the pulsar and stellar winds of PSR B1259-63/SS2883}

The separation distance between the pulsar and optical star in
PSR-B1259-63/LS2883 changes by more than order of magnitude
depending on the orbital phase.  Consequently, the physical
conditions at the interaction point can vary significantly. While
the processes related to particle acceleration and radiation may be
strongly affected by change of the separation distance \citep[see
e.g.][]{khangulyan07}, the regime of hydrodynamical interaction of the
stars depends weakly on this parameter. Indeed, since the
interaction of the winds depends on the ratio of the ram pressures of
the winds,  collision regime remains unaffected by the
change of the separation distance, unless the stellar wind undergoes a
significant acceleration or pulsar wind suffers a strong Compton drag
\citep[see e.g.][]{khangulyan07}. Finally, the interaction at
different orbital phases may differ significantly due to presence of dense
stellar disk in the system.

\subsection{Parameters of the pulsar wind}

\begin{figure}
\centerline{\includegraphics[width=0.5\textwidth,angle=0.0]{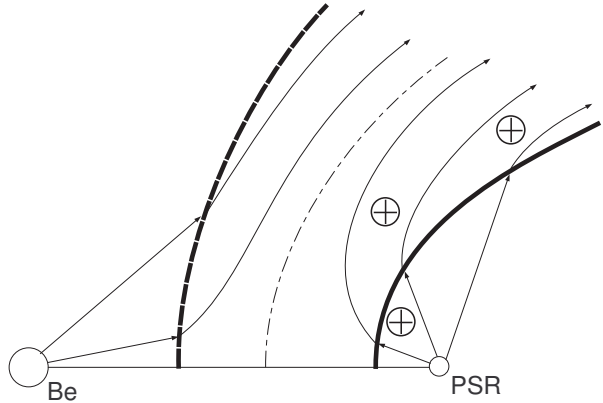}}
\caption{The scheme of the wind interaction and the post shock
  flow. The flow from optical star is terminated by the nonrelativistic
  shock front (thick dashed line). The flow from the pulsar is
  terminated by the relativistic shock front (thick solid
  line). Nonrelativistic and relativistic post shock flows are
  separated by the contact discontinuity (dashed-dotted line). The
  symbol $\oplus$ shows the direction of the toroidal magnetic field
  which has been assumed to exist only in the region occupied by
  pulsar wind.  }
\label{scheme}
\end{figure}

Spin-down luminosity of the pulsar PSR B1259-63 is $L_{sd} = 8 \times
10^{35} {\rm erg\,s^{-1}}$.  This energy is released in the form of
electromagnetic and kinetic energy fluxes carried away by pulsar
wind. Magnetic field in the wind may be represent as superposition of
two components with essentially different spatial dependencies: the
poloidal magnetic field component decreases with radius as $r^{-2}$,
while the toroidal magnetic field has a weaker dependence, namely
proportional to $r^{-1}$. Thus, at distances larger than light
cylinder radius the toroidal magnetic field strongly dominates over
the poloidal one.  Therefore, only the toroidal magnetic field
component has been taken into account in the modelling.  To keep axial
symmetry in the model the axis of rotation of the pulsar is assumed
to be directed along the line connecting the pulsar and optical
star. Figure~\ref{scheme} shows a schematic representation of the wind
interaction geometry.

The magnetic field strength is characterized by the magnetization
parameter $\sigma$, which is the ratio of the magnetic field energy to
kinetic energy fluxes.  Conventional theories of plasma generation in
pulsar magnetospheres predict a highly magnetized wind with $\sigma
\gg 1$ at the light cylinder.  In this case the energy flux from
pulsar is not isotropic and is expected to have the following form
\begin{equation}
\prodi{L}{\Omega} = \gamma_0 mc^2 \prodi{N}{\Omega}+P_0\sin^2(\theta)\,.
\label{eq:flux}
\end{equation}
Here $P_0\sin^2(\theta)$ is the Pointing flux obtained in the
frameworks of the split-monopole model \citep{bogovalov99a};
$\prodi{N}{\Omega}$ is the particle flux in the wind; and $\theta$ is
the angle between the pulsar rotational axis and direction of the
flow.  Obviously, in this case the ratio of the Poynting flux to the
kinetic energy density flux varies with $\theta$.  It is convenient to
define $\sigma$ as the ratio ${P_0/(\gamma_0 mc^2 \prodi{N}{\Omega})}$
which equals to the ratio of the Poynting flux to the kinetic energy
flux in the equatorial plane of the pulsar wind.

Interpretation of observations usually leads to the conclusion that
the wind energy flux at large distances from the pulsar is
concentrated in the particle kinetic energy, i.e. $\sigma\ll1$
\citep{kennel84*b}. Thus, a transformation of the Poynting flux to the
pulsar wind bulk motion is to occur in the pulsar wind zone, although
we note that the physical mechanism responsible for this
transformation still remains unknown.  Assuming that the most of the
Poynting flux is transformed into the kinetic energy flux,
Eq.~(\ref{eq:flux}) can be rewritten as
\begin{equation}
\prodi{L}{\Omega} = \gamma_0 mc^2\prodi{N}{\Omega}+\gamma_{\rm max}\sin^2(\theta)mc^2 \prodi{N}{\Omega} +  P\sin^2(\theta)\,,
\label{eq:flux2}
\end{equation}
where we assumed that the particle flux $\prodi{N}{\Omega}$ is
isotropic. Then the $\sigma$ parameter equals to the ratio
${P/(\gamma_{\rm max}mc^2 \prodi{N}{\Omega})}$. We should note that the
original definition of $\sigma$ in \citet{kennel84*b} does not differ
essentially from the one used here given that $\gamma_{\rm max} \gg
\gamma_0$.  According to this definition $\sigma$ parameter is the 
fraction of the initial Poynting flux remaining in the electromagnetic
form at the wind termination shock.

Thus, the requirement of a kinetically dominated pulsar wind leads to the
conclusion that the wind should be strongly anisotropic and weakly
magnetized.  To study the impact of these effects, below we consider
two different models: (i) weakly magnetized isotropic wind; and (ii)
highly anisotropic unmagnetized wind.

\subsection{Parameters of the stellar wind}
Given the fast rotation of LS2883, the stellar wind is believed to
have a strong angular dependence. Namely, it is expected that the
stellar outflow consists of two parts with comparable mass loss rates:
an almost isotropic polar wind and a dense Keplerian disk in the plane
perpendicular to the star rotation axis. The mass loss rate and
velocity profiles of stellar winds is a subject of future dedicated
studies. We note however that although the precise measurements
presently are not available, the properties of the wind can be
estimated based on the luminosity and temperature of the star
\citep{vink00}. The physical characteristics of the optical star
LS2883 have been recently significantly revised towards higher
temperature and luminosity \citep{negueruela11}. This leads
consequently to a higher mass loss rates. In particular, the mass loss
rate of the polar wind is expected to be at the level of
$\dot{M}=6\times10^{-8} M_\odot \rm yr^{-1}$ \citep{negueruela11},
which exceeds significantly the rate assumed in the previous study by
\citet{bogovalov08}. Thus, the expected ratio of the winds ram
pressures, $\eta$ parameter, for the case of the interaction of the pulsar
wind with the stellar polar wind is as follows:
\begin{equation}
\eta={L_{\rm sd}\over \dot{M}c V_{\rm w}}=5\times10^{-2}\left({\dot{M}\over6\times10^{-8}M_\odot\rm yr^{-1}}\right)^{-1}\left({V_{\rm w}\over 1350\,\rm km\,s^{-1}}\right)^{-1}\,,
\label{eq:eta}
\end{equation}
where the wind velocity has been normalized to the value of the wind termination
velocity $V_{\inf}=1350\pm200\, {\rm km\, s^{-1}}$ \citep{mccollum93}.

Since a rather broad range of variation of the $\eta$ parameter has been
discussed in the previous study by \citet{bogovalov08}, here we adopt two
values of $\eta$: $\eta=0.05$ and $\eta=0.1$, which are
similar to the estimate given by Eq.~(\ref{eq:eta}).

\section{Basic assumptions.}

The evolution of the flow due to the change of the star separation
distance is assumed to be adiabatic, i.e. we consider the collision of
the winds as a steady process. All factors resulting in time-dependent
effects were ignored or intentionally suppressed. In particular, the
tangential instability arising at the interface between the
relativistic and nonrelativistic flows in the post shock region is
suppressed \citep{bogovalov08}. The rotation axis of the pulsar was
assumed to be co-directed with the line connecting the pulsar and the optical
star, thus the flow in the computational domain was considered as
axisymmetric in respect to this line. It implies that all  effects
related to the orbiting of the pulsar are ignored, although these
effects may appear to be very important on larger scales
\citep{bosch-ramon11}.

The Lorentz factor of the particles in the pulsar wind is expected to be
very high, $\gamma_{\rm max} \sim 10^6$ \citep{bogovalov02}. Remarkably,
the post shock flow practically does not depend on $\gamma_{\rm max}$
provided that $\gamma_{\rm max} \gg 1$.  Almost all the energy of the bulk
motion is transformed into  ``thermal'' energy of plasma, thus it
is possible to neglect the rest mass energy of the electrons compare with
the thermal energy and the particle density is not present in the
equations describing the motion of the relativistic pulsar wind in  the post shock
region.

\section{Method of solution}

A special numerical method was used for the integration of the transient HD
(hydrodynamical) and RMHD (relativistic magnetohydrodynamical)
equations \citep{ustyugova99}. The integration of the HD equations was
performed in the region limited by the non-relativistic termination
shock and the contact discontinuity. The RMHD equations were
integrated in the region located between the relativistic termination
shock and the contact discontinuity. The plasma flow outside these two
regions corresponds to a supersonic pulsar and a stellar wind, i.e. it has been 
considered as known. The modelling has been performed with the use of 
an adaptive computational mesh \citep[see for details][]{bogovalov08}.
In the region occupied by the shocked relativistic plasma the RMHD
equations in cylindrical coordinates are
\begin{equation}
\parti{n \gamma}{t} + \frac{1}{r}
\parti{r n \gamma v_r}{r}  + \parti{n \gamma v_z}{z}=0.
\label{rel_first}
\end{equation}
\begin{eqnarray}\nonumber
\parti{\gamma^2 (w+h^2) v_r}{t}+ \frac{1}{r}
\parti{}{r} r \left( (w+h^2) \gamma^2 v_r^2 + (p+\frac{h^2}{2}) \right) + \\
+\parti{}{z} (w+h^2) \gamma^2 v_r v_z  = \frac{p + \frac{h^2}{2}}{r}.
\end{eqnarray}
\begin{eqnarray}\nonumber
\parti{\gamma^2 (w+h^2) v_z}{t}+ \frac{1}{r}
\parti{}{r} r (w+h^2) \gamma^2 v_r v_z + \\
+\parti{}{z} \left( (w+h^2) \gamma^2 v_z^2 + (p+\frac{h^2}{2}) \right) = 0.
\end{eqnarray}
\begin{eqnarray}\nonumber
\parti{ }{t} \left( \gamma^2 (w+h^2) - (p+\frac{h^2}{2}) \right)
+ \frac{1}{r} \parti{}{r} r (w+h^2) \gamma^2 v_r + \\
+\parti{}{z} (w+h^2) \gamma^2 v_z  = 0.
\end{eqnarray}
\begin{equation}
\parti{h \gamma}{t} +
\parti{h \gamma v_r}{r}  + \parti{\gamma v_z h}{z}   =0\ .
\label{rel_last}
\end{equation}
Here $p,~w={4\over 3}p,~n$ and $h$ are the pressure, enthalpy density,
particle density and magnetic field, respectively (all quantities are
in the local rest frame); $v_r$, and $v_z$ are the components of the flow
velocity.

In the region occupied by the shocked nonrelativistic plasma the
equations are
\begin{equation}
\frac{\partial \rho}{\partial t}+ \frac{1}{r} \frac{\partial r
\rho v_r}{\partial r}+ \frac{\partial \rho v_z}{\partial z} =0
\label{eq:nonrel_first}
\end{equation}
\begin{equation}
\frac{\partial \rho v_r}{\partial t} + \frac{1}{r}
\frac{\partial}{\partial r} r \left( \rho v_r^2+p \right) 
+\frac{\partial}{\partial z} \rho v_r v_z = \frac{p}{r}
\end{equation}
\begin{equation}
\frac{\partial \rho v_z}{\partial t} + \frac{1}{r}
\frac{\partial}{\partial r} r \rho v_r v_z +
\frac{\partial}{\partial z} \left( \rho v_z^2+p \right) = 0
\end{equation}
\begin{eqnarray}\nonumber
\frac{\partial}{\partial t} \left( \frac{\rho v^2}{2} + \epsilon
\right) + \frac{1}{r} \frac{\partial}{\partial r} r v_r
\left(\frac{\rho v^2}{2}
+ \epsilon +p \right)+ \\
+\frac{\partial}{\partial z} v_z \left( \frac{\rho v^2}{2} +
\epsilon + p \right) =0
\label{eq:nonrel_last}
\end{eqnarray}
Here $\rho$ and $\epsilon$ are the densities of the mass and the thermal
energy, respectively, and $p=2/3 \epsilon$ is the pressure.  The
computational method used in this work has been developed earlier for
the solution of the equation describing the interaction of a magnetized
relativistic pulsar wind with the interstellar medium \citep[see for
details][]{bogovalov05, bogovalov08}.

\section{Results}

\subsection{Impact of magnetic field}

To investigate the impact of the magnetic field on the process of the
winds collision, a sequence of calculations are performed for
different values of the $\sigma$ parameter and a fixed value of
$\eta=0.1$. The range of considered $\sigma$ parameter has been chosen
to be $\sigma=0.003-0.1$, which corresponds to the conventional
magnetization of the pulsar wind. As mentioned above, the solution in
the hydrodynamical limit ($\sigma=0$) was obtained earlier by
\citet{bogovalov08}.

\begin{figure}
\begin{center}
\subfigure{
\includegraphics[width=.2\textwidth,angle=0.0]{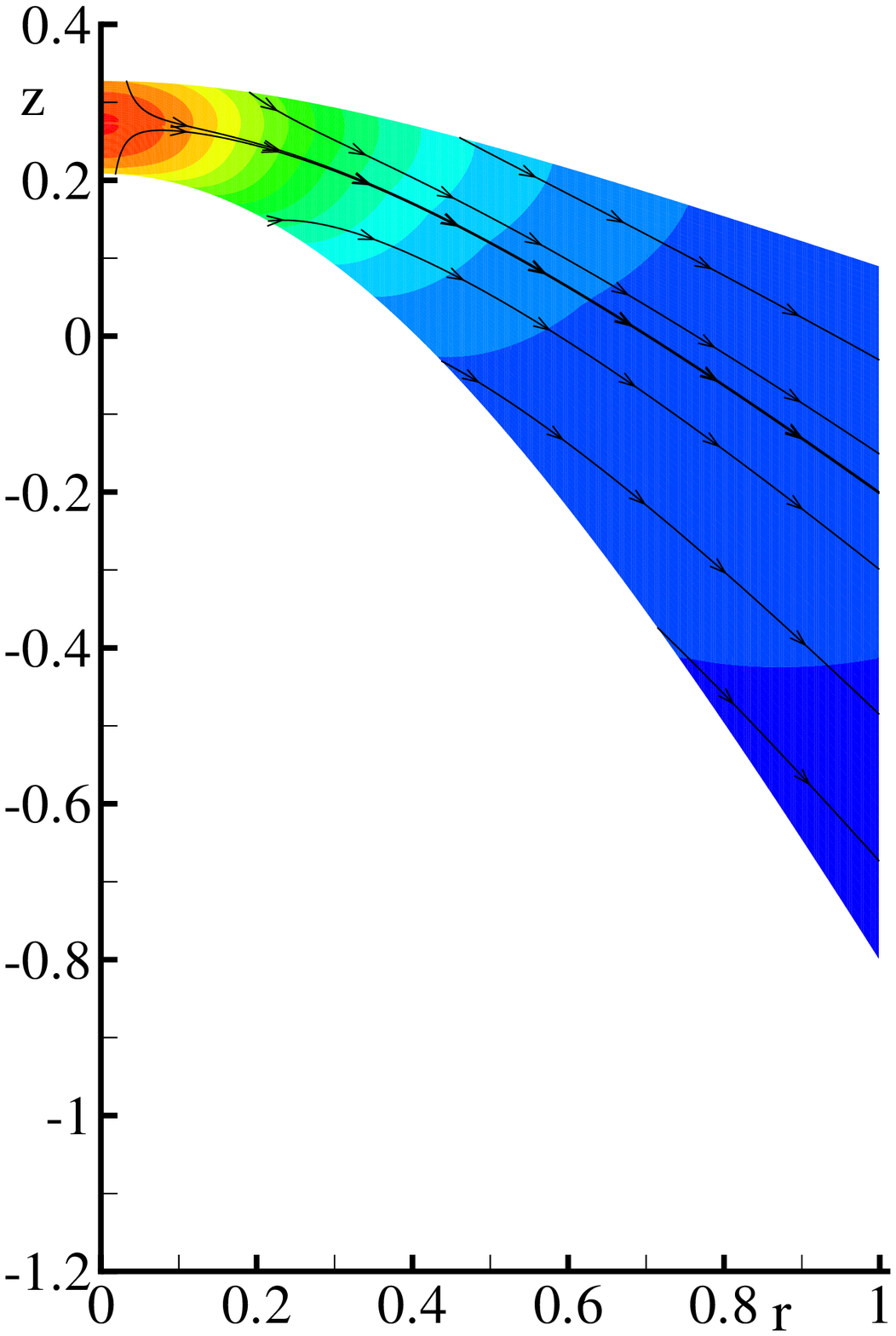}}
\subfigure{
\includegraphics[width=.2\textwidth,angle=0.0]{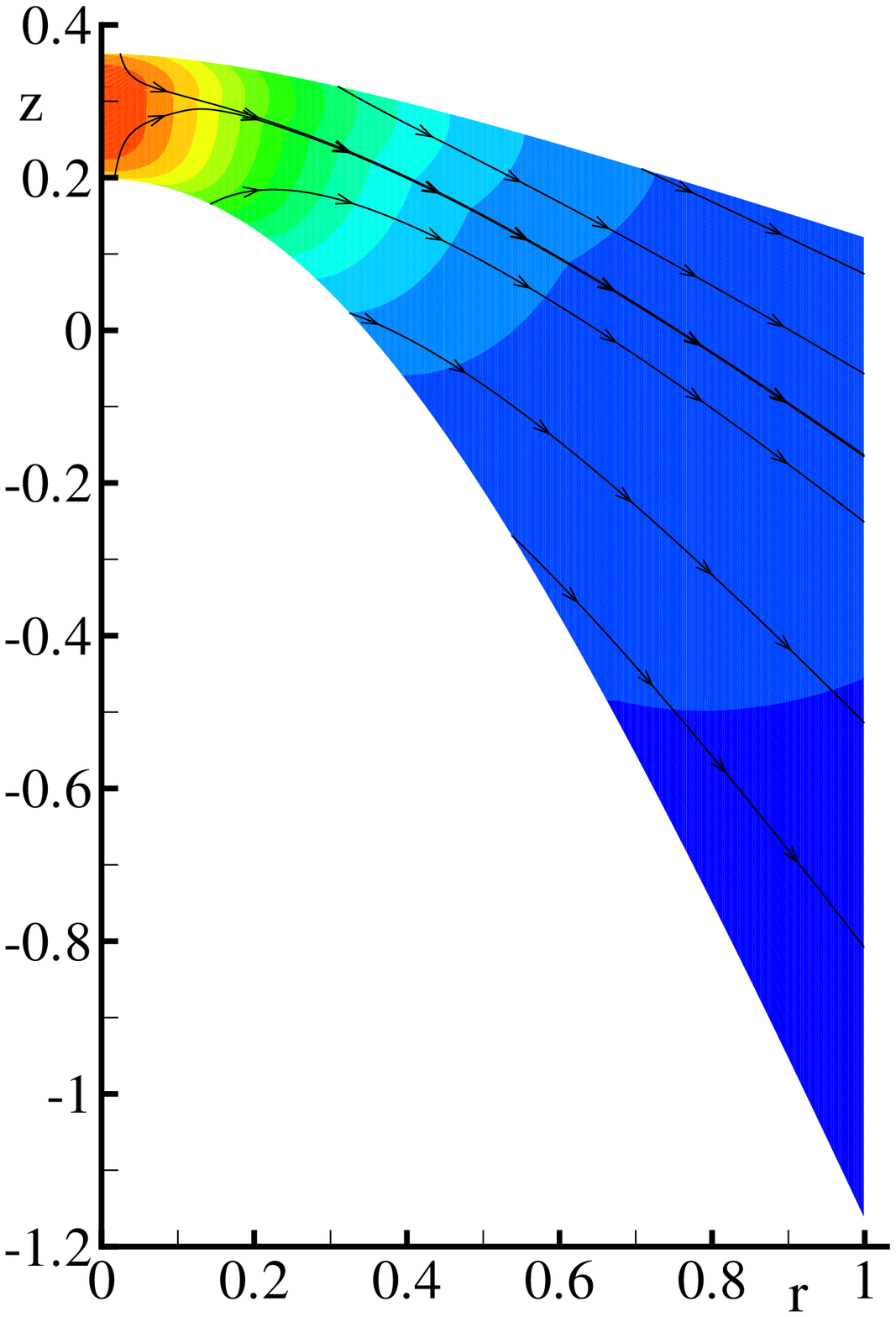}}
\caption{Distribution of total pressure and  stream lines in
  the post shock region for two different values of the magnetization
  of the pulsar wind: $\sigma =0.03$(left panel) and $\sigma=0.1$
  (right panel). The ratio of the winds ram pressures was assumed to
  be $\eta=0.1$.}\label{p1}
\end{center}
\end{figure}

\begin{figure}
\centerline{\includegraphics[width=0.5\textwidth,angle=0.0]{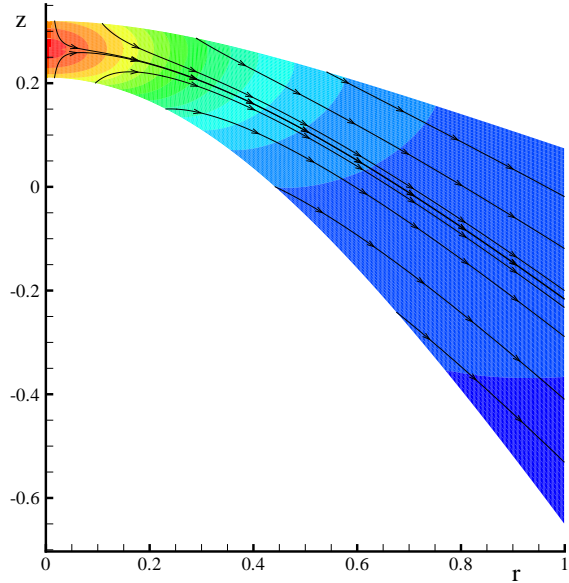}}
\caption{Distribution of total pressure and stream lines in
  the post shock region for the unmagnetized pulsar wind, i.e. $\sigma
  =0$. The ratio of the winds ram pressures was assumed to be
  $\eta=0.1$.}
\label{p0}
\end{figure}

In Figure~\ref{p1} the total pressure distribution (i.e. both gas and
magnetic field pressures $p+h^2/2$) together with stream lines
are shown for different pulsar wind magnetizations. A similar
distribution obtained for the case of an unmagnetized pulsar wind is
shown in Fig.~\ref{p0}. A comparison of these figures shows that the
impact of the magnetic field on the post shock flow is rather
weak. The magnetic field collimates the post shock flow toward the
axis. This happens due to the pinching of the flow by the toroidal
magnetic field \citep[see e.g.][]{bogovalov99}. 
 
\subsection{Anisotropy of the pulsar wind}

As it has been described above, we study the impact of an anisotropic energy
flux in the pulsar wind under the approximation of an unmagnetized
wind. Thus, we have introduced a polar angle dependence of the pulsar wind
bulk Lorentz factor: $\gamma=\gamma_0+\gamma_{\rm max}\sin^2{\theta}$,
where $\gamma_0$ is the initial Lorentz factor of the wind in the
direction of the rotation axis; and $\gamma_{\rm max}$ is the wind Lorentz
factor at the equator. It is expected that the pulsar wind is
strongly anisotropic, i.e. $\gamma_{\rm max} \gg \gamma_0$. The level of
anisotropy can be characterized by the ratio of polar and equatorial
Lorentz factors: $a=\gamma_{\rm max}/ \gamma_0$ . In the calculations we
adopted a value of $a=100$. The definition of the $\eta$ parameter
should be modified in the case of an anisotropic pulsar wind. Namely, it
is natural to assign it to the ram pressure ratio averaged over the polar angle,
i.e. $<\eta>$.

\begin{figure}
\begin{center}
\subfigure{
\includegraphics[width=.2\textwidth,angle=0.0]{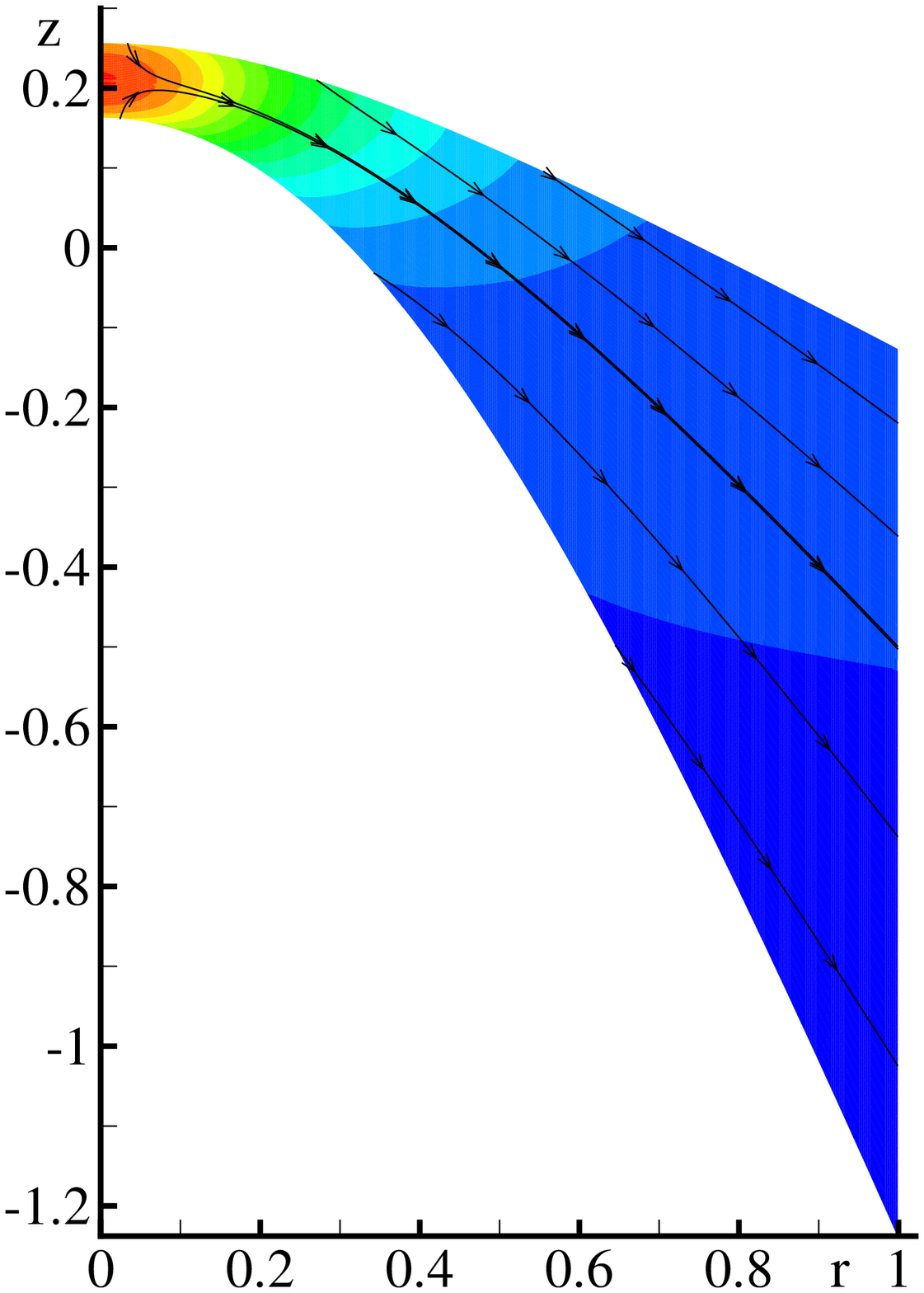}}
\subfigure{
\includegraphics[width=.2\textwidth,angle=0.0]{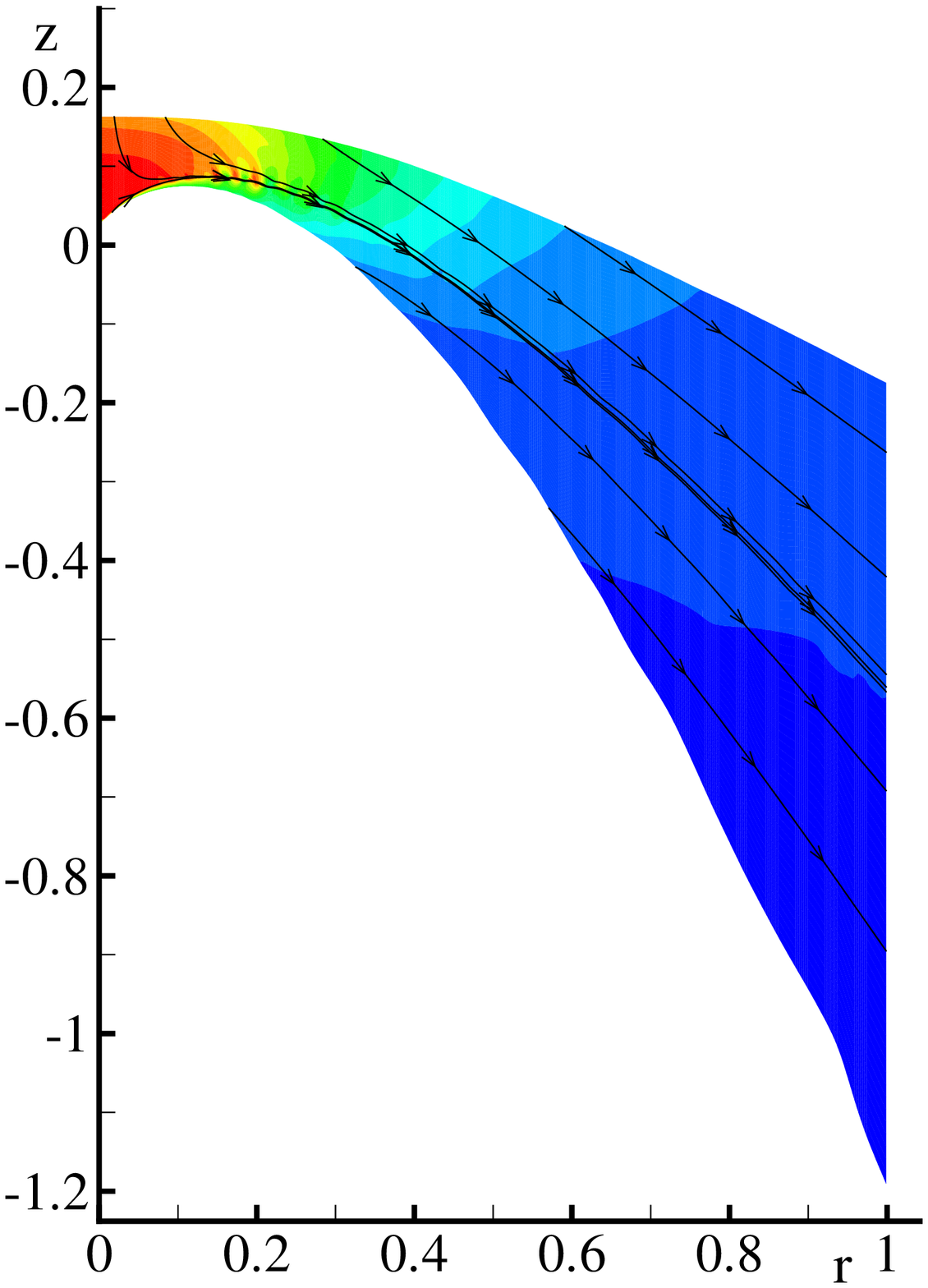}}
\caption{Gas pressure and stream lines in the post shock region for
  the isotropic pulsar wind with $\eta =0.05$ (left panel) and anisotropic
  wind with $<\eta> =0.05$ and $a=100$ (right panel).}\label{piso}
\end{center}
\end{figure}

In Figure~\ref{piso}, a comparison of the pressures and stream lines
is shown for two cases: isotropic pulsar wind with $\eta=0.05$ and
anisotropic pulsar wind with $<\eta>=0.05$ and $a=100$. It might be
seen from the figure that the relativistic terminal shock is located
closer to the pulsar at the axis. This happens because the momentum
flux in this direction is much smaller than the averaged one. It might be
seen as well, that in the case of an anisotropic pulsar wind, the
tangential instability is somewhat more pronounced. In particular,
this is visible on the shape of the pulsar wind termination
shock. Beside these minor changes the geometry of the flow, in the
region down stream farther from the axis, is similar to the flow produced by
an isotropic winds.

\section{Discussion and Conclusions}

\begin{figure}
\begin{center}
\subfigure{
\includegraphics[width=.2\textwidth,angle=0.0]{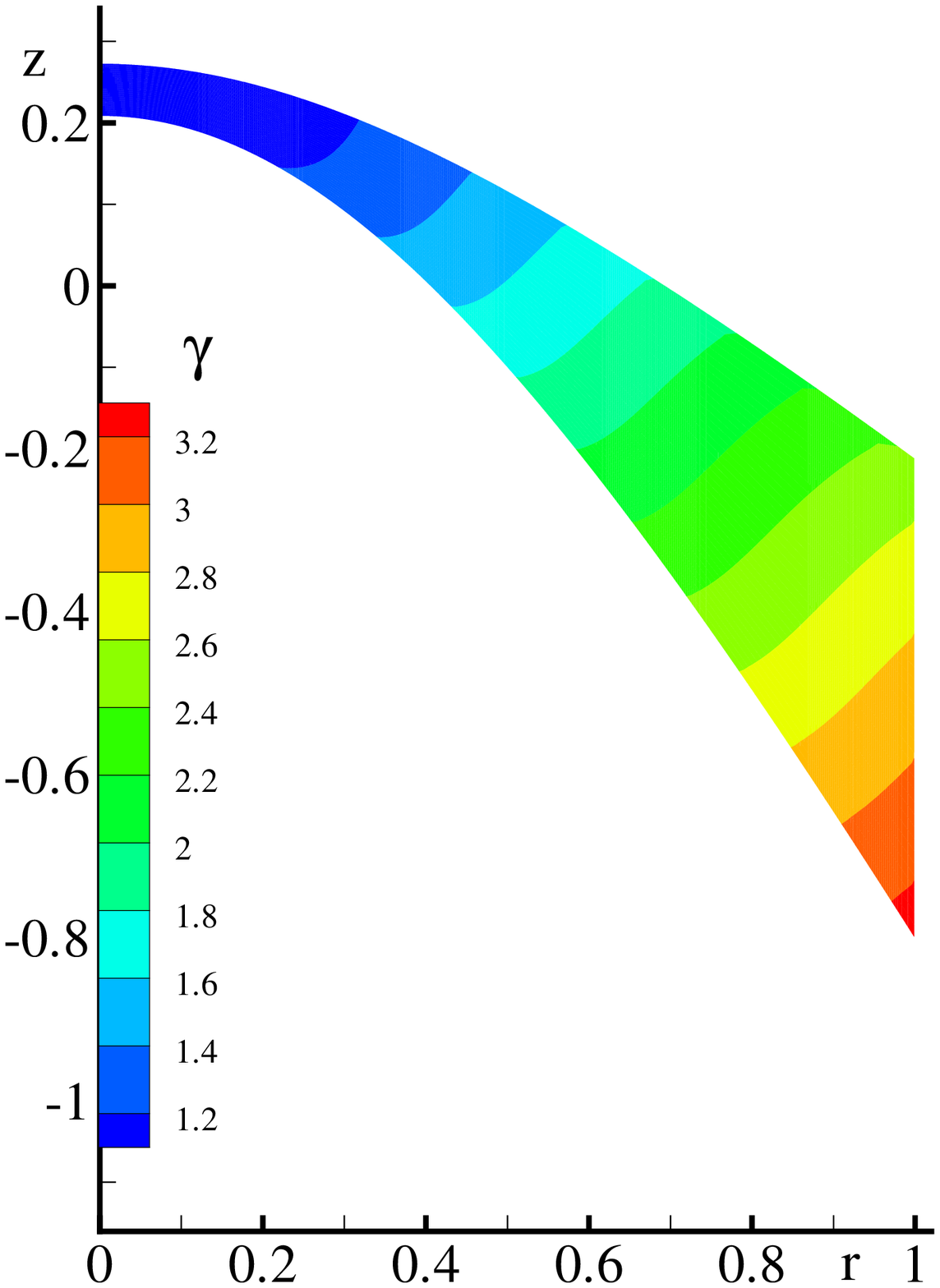}}
\subfigure{
\includegraphics[width=.2\textwidth,angle=0.0]{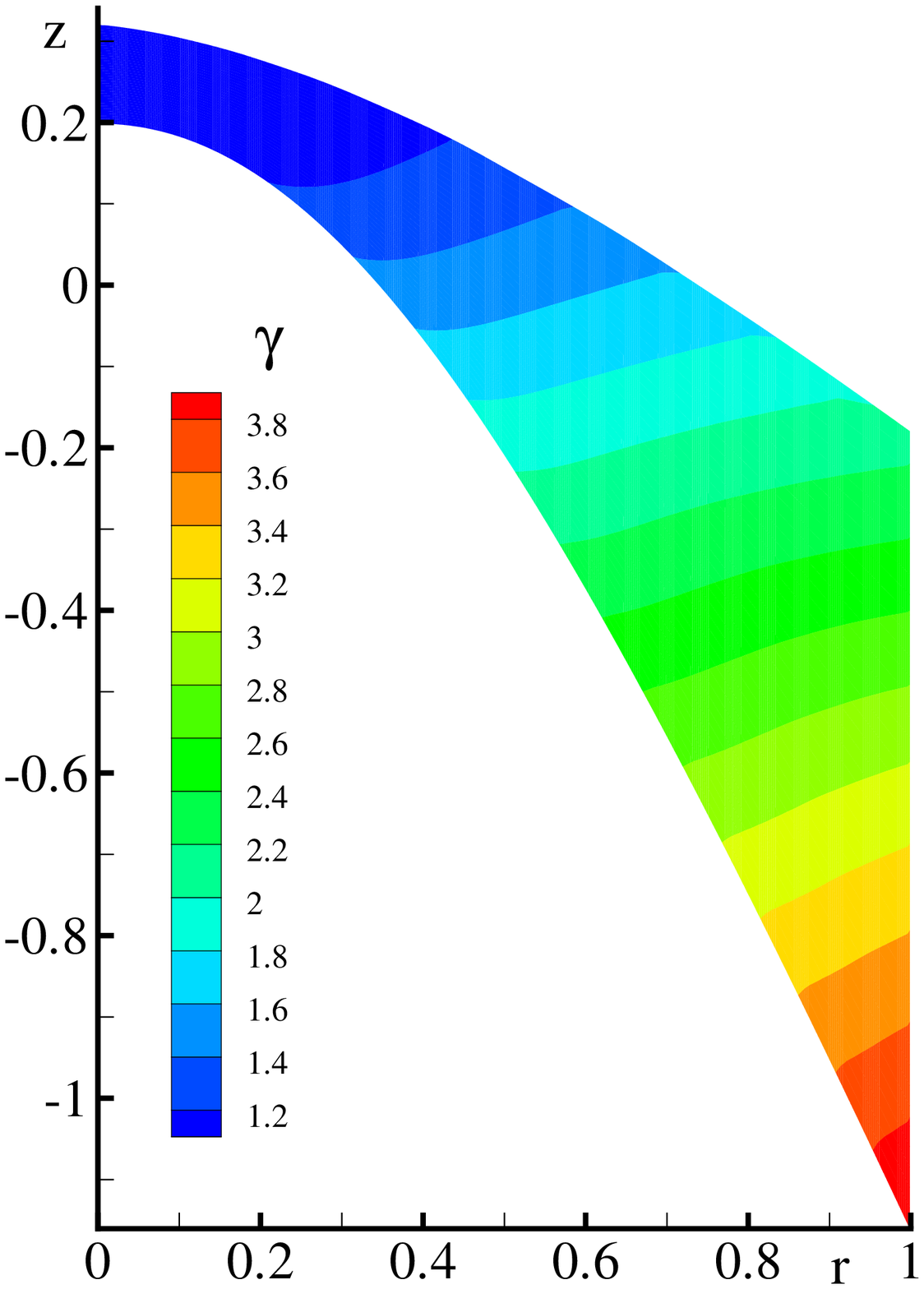}}
\caption{Lorentz factor in the post shock region of the isotropic
  pulsar wind with $\sigma =0.03$ (left panel) and $\sigma=0.1$(right
  panel).}
\end{center}
\label{gamma1}
\end{figure}

The present study of the interaction of a pulsar and a stellar winds
within the framework of an MHD approach shows that the magnetic field
has a rather small impact on the post shock flow. The only apparent
effect revealed by MHD consideration, compared to a pure HD treatment,
is weak collimation of the relativistic flow towards the axis.  This
conclusion is in strong contrast with the case of plerions, i.e. when
the pulsar wind interacts with the interstellar medium
\citep{bucciantini02,khangulyan03,komissarov03,bogovalov05}.  In the
case of plerions even a weak seed toroidal field results in 
efficient collimation of the post shock flow.  This difference has a
rather fundamental reason illustrated in Fig.~\ref{gamma1}, where the
distribution of the bulk motion Lorentz factor is shown.  As it is
seen from this figure, the bulk Lorentz factor is increasing in the
downstream region, i.e. physically the wind collision geometry acts as
a nuzzle, providing an efficient bulk acceleration of the plasma in
the post shock flow. The shock pulsar wind velocity 
fast reaches  relativistic values downstream  the shock. It
implies an inefficient collimation of the plasma, which is a rather
general effect for relativistic outflows \citep{bogovalov99}. In
contrast to this case, if the pulsar wind collides with homogeneous
medium (i.e. interstellar gas), a flow deceleration occurs in the post
shock region.  This results into a few effects making magnetic
collimation to be very efficient \citep{khangulyan03}. The inertia of the
plasma decreases with a decrease of the velocity, thus it is easier to
change the direction of the flow even by a small force. Additionally, since in this case
 the flow is subsonic, the toroidal magnetic field increases with the
distance from the shock. For this reason even a weak initial field is
amplified quickly and can provide an efficient collimation of the flow
towards the axis.

The impact of the pulsar wind  magnetization and anisotropy 
is investigated for case of the collision of the pulsar wind with
the non-relativistic wind from a massive star.  It is shown, that the magnetic
field leads to weak collimation of the post shock flow toward the
axis. The effect remains very weak for a broad parameter range, in
particular for the strongest considered magnetization of $\sigma =
0.1$.  This weak impact of the magnetic field on the wind collision is
the result of the acceleration of the plasma bulk motion due to  adiabatic
cooling. This implies that the calculation of
the radiation produced at the wind collision can be performed in the
hydrodynamical limit in which the toroidal magnetic field is
calculated as a passive scalar not affecting dynamics of the plasma.

The influence of anisotropy of the pulsar wind appeared as well to be
small. Whereas its impact can be seen on the shape of the pulsar wind
termination shock and the enhanced potential rate of instabilities, the general
structure of the flow remains almost identical to the case of
isotropic pulsar wind. Although, we note that in the case of a different
orientation of the pulsar equatorial plane, a stronger effect can
arise.

\section*{Acknowledgments}
The authors are grateful to V.~Bosch-Ramon for his interest to this
study and useful discussions.  The work of S.V.Bogovalov have been
supported by the Federal Targeted Program "The Scientific and
Pedagogical Personnel of the Innovative Russia" in 2009-2013 (the
state contract N 536 on May 17, 2010). The work of A.V.Koldoba and
G.V.Ustugova has been supported by RFBR grants Nr.~09-01-00640a and
Nr.~09-02-00502a


\end{document}